\PassOptionsToPackage{unicode}{hyperref}
\PassOptionsToPackage{hyphens}{url}
\documentclass[
]{article}

\usepackage{amsmath,amssymb,mathabx,commath}
\usepackage{float}

\usepackage{tikz}
\usetikzlibrary{fit, positioning, shapes.geometric, decorations.pathreplacing, calc, arrows.meta, positioning}
\usepackage{tikz-cd}

\tikzset{stretch/.initial=1}
\newcommand\drawloop[4][]%
{\draw[shorten <=0pt, shorten >=0pt,#1]
($(#2)!\pgfkeysvalueof{/tikz/stretch}!(#2.#3)$)
let \p1=($(#2.center)!\pgfkeysvalueof{/tikz/stretch}!(#2.north)-(#2)$),
    \n1= {veclen(\x1,\y1)*sin(0.5*(#4-#3))/sin(0.5*(180-#4+#3))}
    in arc [start angle={#3-90}, end angle={#4+90}, radius=\n1]%
}

\usepackage[margin=1.5in]{geometry}

\makeatletter
\@ifundefined{KOMAClassName}{
    \IfFileExists{parskip.sty}{%
        \usepackage{parskip}
    }{
        \setlength{\parindent}{0pt}
        \setlength{\parskip}{6pt plus 2pt minus 1pt}}
}{
    \KOMAoptions{parskip=half}}
\makeatother

\usepackage{longtable,booktabs,array}
\usepackage{calc} 
\usepackage{subcaption}
\usepackage{graphicx}
\makeatletter
\def\fps@figure{htbp}
\makeatother
\usepackage{adjustbox}

\makeatother

\usepackage{hyperref}
\hypersetup{
    pdftitle={Structure \& Quality},
    pdfauthor={Ryan Williams},
    hidelinks}

\usepackage{url}
\usepackage{natbib}
\usepackage{mathtools}
\usepackage{amsfonts}

\DeclareMathOperator*{\expectation}{\mathbb{E}}
\DeclareMathOperator*{\Var}{\mathrm{Var}}

\newcommand{\appropto}{\mathrel{\vcenter{
    \offinterlineskip\halign{\hfil$##$\cr
    \propto\cr\noalign{\kern2pt}\sim\cr\noalign{\kern-2pt}}}}}

\title{Structure \& Quality}
\makeatletter
\providecommand{\subtitle}[1]{
    \apptocmd{\@title}{\par {\large #1 \par}}{}{}
}
\makeatother
\subtitle{Conceptual and Formal Foundations for the Mind-Body Problem}
\author{ Ryan
Williams\vspace{0.05in} \\ \newline\normalsize\url{ryan@cognitivemechanics.org} }
\date{April 2025}

\begin{document}

    \newgeometry{margin=1in}
    \thispagestyle{empty}
    \pagenumbering{gobble}

    \begin{center}
    {\Large\bfseries One-Page Summary: Structure \& Quality\\[4pt]
    \normalsize Conceptual and Formal Foundations for the Mind-Body Problem}\\[6pt]

    \vspace{10pt}

    Ryan Williams\\
    \texttt{ryan@cognitivemechanics.org}
    \end{center}

    \vspace{10pt}

    \thispagestyle{empty}

    \textbf{Overview.} This article reframes the mind–body problem by analyzing the relationship between structural and
    qualitative properties of a system.
    By formalizing five canonical models of structure–quality interaction, the paper introduces a unified 2D geometric
    representation called \emph{Q–S space}, which enables quantification of mutual determinability between the domains.
    Coordinates in this space have been applied in a companion paper \emph{Qualia \& Natural Selection} to derive
    formal constraints on the evolution of consciousness.

    \vspace{10pt}

    \textbf{Five Models.} Five models of structural-qualitative relationship are presented:
    \begin{enumerate}
        \item \textbf{Functionalism}: Evolves according to structure only.
        \item \textbf{Structural–Determination}: Qualities are epiphenomenal but derived from structure.
        \item \textbf{Qualitative–Determination}: Structural dynamics are wholly grounded in qualitative states.
        \item \textbf{Mixed–Determination}: Structure and quality are both efficacious.
        \item \textbf{Equivalent Systems}: Dual structural and qualitative descriptions yield identical dynamics.
    \end{enumerate}

    \vspace{10pt}

    \textbf{Conceptual Illustrations.} Toy models of each illustration is presented as a 2D cellular automaton.

    \begin{figure}[H]
        \centering
        \scriptsize
        \begin{subfigure}[t]{0.2\textwidth}
            \centering
            \includegraphics[width=\textwidth]{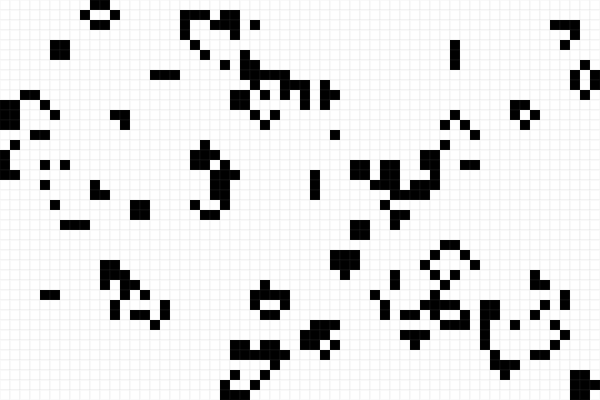}
            \caption*{\scriptsize Bare Functionalism}
        \end{subfigure}
        \begin{subfigure}[t]{0.2\textwidth}
            \centering
            \includegraphics[width=\textwidth]{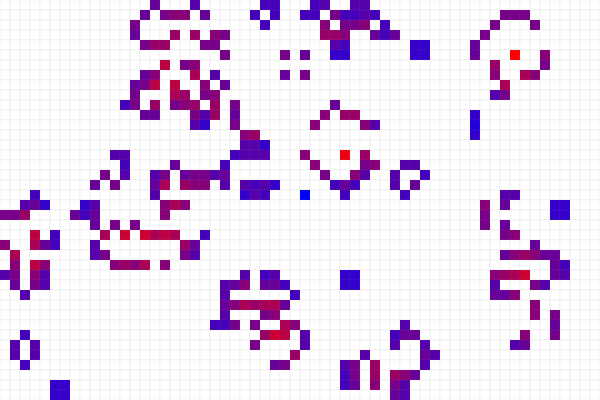}
            \caption*{\scriptsize Structurally-Determined}
        \end{subfigure}
        \begin{subfigure}[t]{0.2\textwidth}
            \centering
            \includegraphics[width=\textwidth]{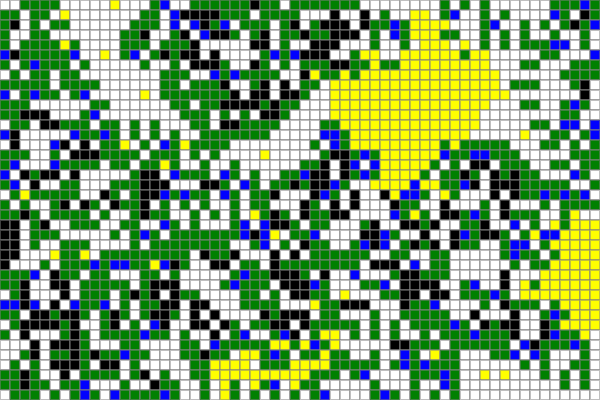}
            \caption*{\scriptsize Qualitatively-Determined}
        \end{subfigure}
        \begin{subfigure}[t]{0.2\textwidth}
            \centering
            \includegraphics[width=\textwidth]{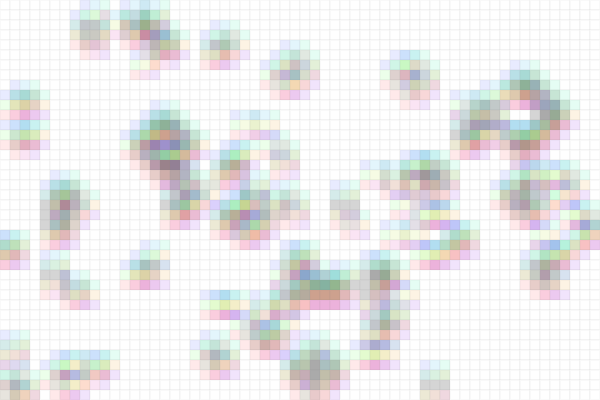}
            \caption*{\scriptsize Equivalently-Determined}
        \end{subfigure}
        \label{fig:ca-illustrations-1}
    \end{figure}

    \textbf{Q--S Space.} A coordinate pair \( (\hat{q}, \hat{s}) \) represents the degree to which
    information about structure determines quality and vice versa.
    For example, $(\hat{q}=.5, \hat{s}=1)$ represents a structurally-determined system; $q$ is wholly determined by $s$,
    but uncertainty about $s$ is only resolved 50\% by knowledge of $q$.

    \begin{figure}[htbp]
        \centering
        \begin{tikzpicture}[scale=2.5, font=\scriptsize]

            \draw[->] (0,0) -- (1.1,0);
            \draw[->] (0,0) -- (0,1.1);

            \node[rotate=0, anchor=north, font=\small] at (0.5, -0.02) {$\hat{q}$};
            \node[rotate=90, anchor=south, font=\small] at (-0.02, 0.5) {$\hat{s}$};

            \draw[dashed, gray] (1,0) -- (1,1);
            \draw[dashed, gray] (0,1) -- (1,1);

            \node[circle, draw, inner sep=1.2pt, line width=1pt, label=below left:{\textbf{Functional} (0,0)}] at (0,0) {};
            \node[circle, draw, inner sep=1.2pt, line width=1pt, label=above left:{\textbf{Structural} ($\hat{q}$,1)}] at (0,1) {};
            \node[circle, draw, inner sep=1.2pt, line width=1pt, label=below right:{(1,$\hat{s}$) \textbf{Qualitative}}] at (1,0) {};
            \node[circle, draw, inner sep=1.2pt, line width=1pt, label=above right:{(1,1) \textbf{Equivalent}}] at (1,1) {};
            \node[circle, draw, inner sep=1.2pt, line width=1pt, label=above:{\textbf{Mixed}}] at (0.5,0.5) {};

        \end{tikzpicture}
        \label{fig:entropy-map-1}
    \end{figure}

    \textbf{Interpretation.} The article explores interpretations of each of the five models within the conventional
    taxonomy of functionalism, dualism, emergent materialism, panpsychism, idealism, and neutral monism.

    \vspace{10pt}

    \textbf{Implications.} Serves as the foundation for a quantitative theory of qualitative evolution developed in the companion paper \emph{Qualia \& Natural Selection}.

    \restoregeometry 
    \clearpage
    \pagenumbering{arabic}  
    \newpage

    \maketitle
    \begin{abstract}
        This paper explores the hard problem of consciousness from a different perspective.
        Instead of drawing distinctions between the physical and the mental, an exploration of a more foundational relationship is examined:
        the relationship between structure and quality.
        Information-theoretic measures are developed to quantify the mutual
        determinability between structure and quality, including a novel \emph{Q–S space} for analyzing fidelity between the two domains.
        This novel space naturally points toward a five-fold categorization of possible relationships between structural and qualitative
        properties, illustrating each through conceptual and formal models.
        The ontological implications of each category are examined, shedding light on debates around functionalism,
        emergentism, idealism, panpsychism, and neutral monism.
        This new line of inquiry has established a framework for deriving theoretical constraints on qualitative systems
        undergoing evolution that is explored in my companion paper, \emph{Qualia \& Natural Selection}.\footnote{\cite{williams2025qualia}}
    \end{abstract}

    This paper takes a new perspective of the hard problem of consciousness by changing the fundamental entities that are
    examined.
    Instead of trying to revisit difficult questions about the relationship of the body and mind directly, we instead
    take a more basic perspective by exploring the possible relationships between structure and quality.

    In our investigation, we explore toy models developed in cellular automata systems that give intuition for the possible
    relationships between structural and qualitative domains.
    Those intuitions point us to a five-fold categorization of structure-quality relationships, which then leads us to
    formalize the dynamics and informatics of such systems.

    The end result of the inquiry is the classification of structure-quality relationships via their informational structure.
    These architectures are then plotted within a 2-D informational space called \emph{Q-S space}, which lays out geometrically
    how much information about the structural domain determines features of the qualitative domain, and vice versa.

    This new \emph{Q-S space} is put into use in my companion paper, \emph{Qualia \& Natural Selection}, which
    derives statistical bounds on quantities known to evolutionary theory via the Price Equation,
    opening a new landscape for a quantitative investigation of natural selection and its efficacy in the qualitative domain.

    \subsection{Background}\label{subsec:background}

    In this first section, we will provide some basic philosophical background about what we mean by the terms
    \emph{structure} and \emph{quality}, and give a basic outline of the five-fold categorization of the potential types
    of relation between structure and quality.

    \subsubsection{Structure \& Quality}\label{subsubsec:structure-quality}

    The move to the more basic notions of structure and quality is the key insight that opened the opportunity for
    a formal inquiry into the relationship of physical and mental phenomena.
    A brief pause to clarify the meaning of these terms as used in this paper is warranted.

    Structure refers to aspects of the world that are amenable to being described completely in abstract terms.
    Other terms for this notion include quantitative or relational.
    When we describe the world with equations we are using a structural description.
    The structural description of the world discards any intrinsic properties, instead classifying the state of the world
    according to the abstract relationships that those states contain.

    Quality refers to the ineffable, intrinsic properties that we find through the window of our experience.
    I generally opt for the term ``qualitative properties'' because it is a more neutral term, but the notion includes
    what would often be described as qualia.
    We will see as we get into the formal descriptions of qualitative-structural systems below that
    the qualitative domain may also exhibit its own structure.
    This structure is necessary for there to be any map between the qualitative and structural domains.

    But the important feature of the qualitative is that it does not pretend to be an abstract, unanchored domain.
    It is entirely grounded in the qualitative properties of the world, of which we are familiar from our own conscious
    experience.

    \subsubsection{Relationships Between Structure \& Quality}\label{subsubsec:relationships}

    At bottom analysis, there appear to be five general categories of relationship between structural and qualitative properties.
    Each relationship type offers a different framework for interpreting how qualitative and structural features might correlate.

    \begin{enumerate}
     \item We will call the first relationship \emph{bare functionalism} or just \emph{functional}.
    This relationship denies either the existence or relevance of qualitative properties.
    The world can be accounted for entirely via structural properties, without any fundamental qualitative aspect.

    \item The second relationship will be referred to as \emph{structurally-determined} or just \emph{structural}.
    In a structurally-determined system, qualitative properties are determined by structural properties.
    Imagine a world where the faster-moving particles are more red and the slower ones are more blue; their speed determines their color.

     \item The third relationship is called \emph{qualitatively-determined} or \emph{qualitative}.
     A system that is \emph{qualitatively-determined} means that its structural behavior is defined in terms of the qualitative properties of its
     constituents.
     A qualitatively-determined system is one whose states can be defined entirely in terms of sets of qualities.
     This tends to give rise to rather abstract systems, an example of which will be illustrated below.

     \item The fourth relationship we will call \emph{mixed-determination} or just \emph{mixed}.
     Picture a world where the red particles go up, and the blue particles go down; it is part of their intrinsic nature
     as red and blue particles that they behave structurally as they do.
     \footnote{In this paper, we will generally use the convention that different colors indicate different qualities,
         and that some spatial arrangement indicates the structural, functional, relational, and/or quantitative aspects of the system.}
    In practice, the behavior of most qualitatively-determined systems is partly-determined by structure and partly determined by quality.

    \item The fifth relationship is an \emph{equivalently-determined} or just \emph{equivalent} system.
    In an equivalent system the same behavior can be described in intrinsic (qualitatively-determined or mixed-determination)
     terms and equivalently in structural terms.
    There is no preferred description as they yield the same behavior.
    This approach may have the potential to unite positive aspects of both structural and intrinsic systems, as we will see below.
    \end{enumerate}

    Intuitively, this set of five relationships seems to constitute an exhaustive categorization of the possible types of
    relationship between quality and structure, but I do not claim it as a deductive certainty.

    None of these relations between structure and quality is to be taken as necessary.
    We do not present (nor believe that there is) any way to derive qualities out of structure from first principles.
    Ultimately, the relationships are simply implied to be descriptive accounts of the world as it might be, as any
    other postulated ontological entities within a scientific theory.

    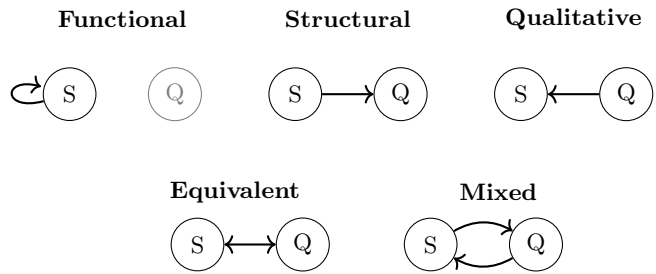
\begin{figure}[htb]
        \centering
        \begin{tikzpicture}[node distance=1cm and 1.5cm, auto, font=\small]

            \tikzstyle{circlebox} = [circle, draw, minimum size=.7cm, align=center]

            \node (f_title) at (0,2) {\textbf{Functional}};
            \node[circlebox] (s1) at (-0.7,1) {S};
            \node[circlebox, gray] (q1) at (0.7,1) {Q};
            \draw[->, thick] (s1) edge[loop left] (s1);

            \node (st_title) at (3,2) {\textbf{Structural}};
            \node[circlebox] (q2) at (2.3,1) {S};
            \node[circlebox] (s2) at (3.7,1) {Q};
            \draw[->, thick] (q2) -- (s2);

            \node (i_title) at (6,2)  {\textbf{Qualitative}};
            \node[circlebox] (s3) at (5.3,1) {S};
            \node[circlebox] (q3) at (6.7,1) {Q};
            \draw[<-, thick] (s3) -- (q3);

            \node (e_title) at (1.5, -0.3) {\textbf{Equivalent}};
            \node[circlebox] (s4) at (1, -1) {S};
            \node[circlebox] (q4) at (2.4, -1) {Q};
            \draw[<->, thick] (s4) -- (q4);

            \node (m_title) at (5, -0.3) {\textbf{Mixed}};
            \node[circlebox] (s5) at (4.1, -1) {S};
            \node[circlebox] (q5) at (5.5, -1) {Q};
            \draw[->, thick] (s5) edge[bend left] (q5);
            \draw[->, thick] (q5) edge[bend left] (s5);

        \end{tikzpicture}
        \caption{Causal dependency models for five types of structure–quality relationships.
        Here we understand ``cause'' in the fairly loose sense that the ``effects'' can be derived from the causes,
            though not necessarily so in the opposite direction.
            The double arrow of the equivalent structure can be seen to mean that each aspect determines the other.
            The curved arrows of the mixed system indicate that each partly determines the other, but neither is a full
            description of the system.}
        \label{fig:causal-loops}
    \end{figure}

    \subsection{Conceptual Illustrations}\label{subsec:conceptual-illustrations}

    In this section, conceptual illustrations are given of each of the five categories of relationship between
    structure and quality.
    The goal is to give an intuitive feel for how systems within each of these categories might operate in the setting of
    toy models composed of 2D cellular automata.\footnote{\cite{vonneumann1966}, \cite{wolfram2002}}

    \subsubsection{Bare Functionalism}\label{subsubsec:functionalism-illustration}

    A bare functionalist system is one that only acknowledges structural rules.
    It has no built-in qualitative properties.
    Certain philosophical positions hold a bare functionalist view of the world, in which no qualitative aspects
    are included in their model of the world.
    This includes worldviews in which qualitative aspects of the world are illusory.

    Though I am skeptical about the power of this type of model as an explanation of the world, it is
    a common position and should be illustrated as a baseline for what a purely structural description looks like.

    As will be a theme in this section, we will consider 2D cellular automata as illustrations of the various types of models of the
    world.
    This case is actually the simplest, because it involves no introduction of anything new.
    We can simply show any traditional cellular automata, such as Conway's Game of Life.\footnote{\cite{Gardner1970}}

    In other models, we will represent qualitative properties via the addition of colors to the automata.
    The traditional black/white (representing 1/0) of the CA show how this picture lacks any place for the inclusion of qualitative properties
    as part of the model.
    There is only pure structure.

    \subsubsection{Structurally-Determined Systems}\label{subsubsec:structural-illustration}

    In a structurally-determined system, the structural behavior of the system can be described entirely without qualities.
    The unfolding of the rules is unaffected by them.
    However, in contrast to bare functionalism, structurally-determined qualities still incorporate qualitative properties
    as part of the model.
    The qualitative properties are derived by rules that can be determined entirely from the system's structure; but the
    qualitative properties are still there.

    We can illustrate this picture with a CA that obeys Conway's Game of Life, but also has additional rules for how
    to assign colors (qualities) to the structure.
    In the example illustration, the color of each cell is determined by how many neighbors it has concentrated in its
    general vicinity; more neighbors equates to a redder color and fewer equates to a bluer tone.\footnote{Inspired by an example from \cite{hume1739}.}

    Note that in this picture, if we were to turn it black and white, we would have no need for the qualities to determine
    which of the cells would be off or on in future steps.
    We could evolve it forward without any knowledge of the qualities.
    This is the key distinguishing feature of a structural system.

    \subsubsection{Qualitatively-Determined Systems}\label{subsubsec:qualitatively-determined-illustration}

    A qualitatively-determined system is one whose behavior can be determined entirely by knowing what set of qualities
    it holds.
    Imagine an unstructured set of possible colors; each combination of the colors represents a unique state.
    So the state \emph{red-green}, is distinct from \emph{red-blue}, which are both distinct from \emph{red-green-blue}, etc.

    \begin{figure}[htbp]
        \centering
        \begin{subfigure}[b]{0.48\textwidth}
            \centering
            \includegraphics[width=\textwidth]{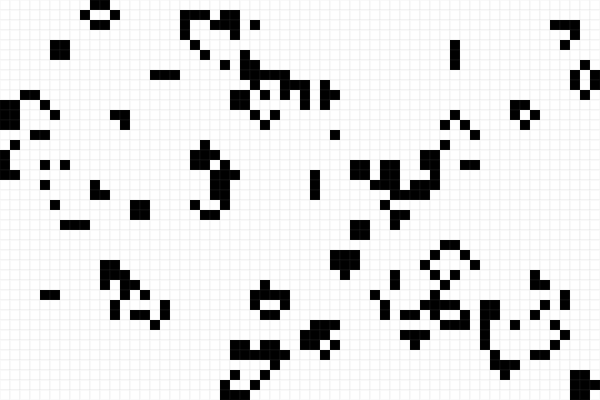}
            \caption{Bare Functionalism}
            \label{fig:bare-functionalism}
        \end{subfigure}\hfill
        \begin{subfigure}[b]{0.48\textwidth}
            \centering
            \includegraphics[width=\textwidth]{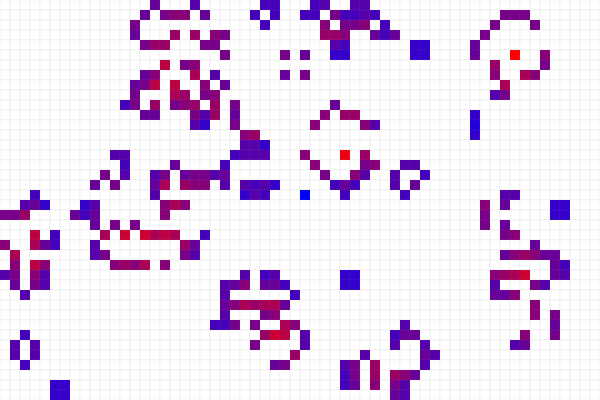}
            \caption{Structurally-Determined System}
            \label{fig:structurally-determined}
        \end{subfigure}

        \vspace{0.5cm}

        \begin{subfigure}[b]{0.48\textwidth}
            \centering
            \includegraphics[width=\textwidth]{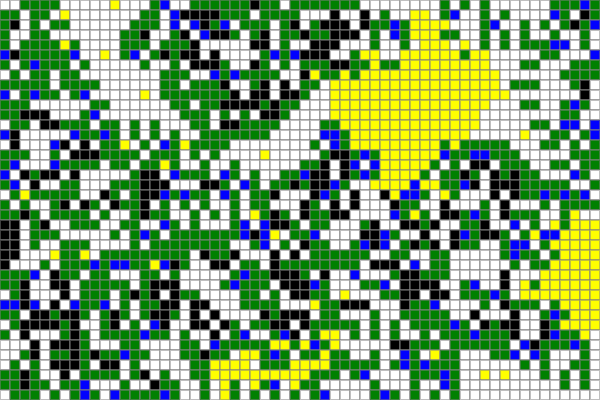}
            \caption{Qualitatively-Determined System}
            \label{fig:qualitatively-determined}
        \end{subfigure}\hfill
        \begin{subfigure}[b]{0.48\textwidth}
            \centering
            \includegraphics[width=\textwidth]{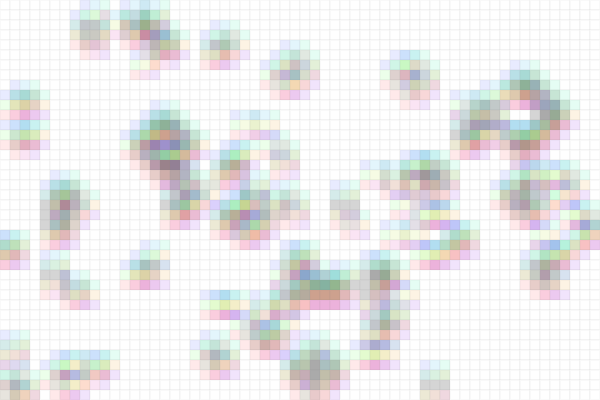}
            \caption{Equivalently-Determined System}
            \label{fig:equivalently-determined}
        \end{subfigure}

        \caption{
            Illustrations of four categories of relationships between structure and quality, using 2D cellular automata examples.\\
            \\
            (\subref{fig:bare-functionalism}) A purely functional system, where states are structural only, lacking intrinsic qualitative properties.\\
            \\
            (\subref{fig:structurally-determined}) A structurally-determined system, where qualitative properties (color) depend solely on structural configurations.\\
            \\
            (\subref{fig:qualitatively-determined}) A qualitatively-determined system, whose evolution explicitly depends on qualitative states.\\
            \\
            (\subref{fig:equivalently-determined}) An equivalently-determined system, whose behavior admits both structural and qualitative descriptions, each fully capturing the system dynamics.
            Note that the color of each cell in the equivalent system encodes the underlying active and inactive structural states; certain colored cells actually indicate an inactive cell.
        }
        \label{fig:ca-illustrations}
    \end{figure}

    Each state can be fully understood in terms of which colors are present, and we make no presumption of how many
    colors there are, or even whether they are continuous or discrete.
    We can imagine transition rules like $RG \to RGB$, $RGB \to RB$, etc., which constitute a full description of
    the behavior of the system.

    Now imagine that some subset of these colors, say any hue with some blue in it, are projected into an ordered
    1-dimensional space (a spectrum of sorts) of different colors of blue.
    The combination of colors with blue wavelengths from our state are projected onto this space and considered in some
    way ``active''.

    We can see that this system has a structural representation as active points on the blue spectrum which is
    fully-determined by the unstructured set of colors in the qualitative state.
    However, since the behavior of the qualitative state contains colors outside of the blue spectrum, and since the
    dynamics might directly consider those other colors in the qualitative state, the structural representation will
    only be a partial picture, both of the current states and to understand the dynamics.

    \subsubsection{Mixed-Determination Systems}\label{subsubsec:mixed-illustration}

    In a mixed-determination system, the structural behavior of the system cannot be fully described without reference
    to its qualities.
    We can think about this as a system whose rules are defined in terms of the qualities of each cell.
    So for instance, a red cell has different behavior from a blue cell.
    The red cell might die unless it has exactly two neighbors of any color, but the blue cell might require at least
    two specifically blue neighbors to survive, and so on.

    Notice that we cannot evolve these systems forward without knowing which color each cell is.
    The quality is a critical part of applying the rules themselves.

    One thing to keep in mind is that the example above is of a \emph{mixed-determination} system; its evolution is
    governed by its structure and its quality, but cannot be fully determined by either domain alone.

    \subsubsection{Equivalently-Determined Systems}\label{subsubsec:equivalent-illustration}

    An equivalently-determined system is one which has an qualitative description and a structural description
    which both describe the same behavior of the system.
    The qualitative description may be fully qualitatively-determined or mixed-determination.

    The example from our illustration utilizes a system which evolves functionally according to Conway's Game of Life.
    The color of each cell encodes information about exactly which of the 9 cells
    (including itself) in its immediate vicinity are active (some colors are actually ``inactive'').
    This allows the system to be viewed from one perspective as a grid of active and inactive cells, determining
    each cell's colors from its state along with that of each of its neighbors.
    From an qualitative perspective, the rules can be reframed in terms of the color of each cell.
    A cell with one color will behave differently from a cell of another color.
    The same structural dynamics can be recovered from the mixed-determination representation.

    This specific example will be explored in more detail below, along with another example that uses a qualitatively-determined
    representation.

    \subsection{Dynamics}\label{subsec:dynamics}

    Now that we've demonstrated each of the relationships conceptually, we can now give more precise
    formulations of the dynamics of these systems.

    \subsubsection{Functionalist Dynamics}\label{subsubsec:functionalist-dynamics}

    In a bare functionalist system, the dynamics can be completely specified by a set of relations $F$ where each $F_i$
    can be seen as a map from the set of possible (structural) states $S$, and time $T$, defined in relation to the
    time described by the associated state.
    \begin{equation}\label{eq:structural-dynamics}
    F_i : S \times T \to S
    \end{equation}
    Note that $T$ may be $\mathbb{R}$ in the case of continuous-time systems and $\mathbb{Z}$ in the case of
    discrete-time systems; either choice will not affect our present discussion.

    \subsubsection{Structural Dynamics}\label{subsubsec:structural-dynamics}

    In a system with structurally-determined qualities, we take the dynamics to evolve structurally according to
    (\ref{eq:structural-dynamics}).
    We also require an additional mapping $\phi$ from the set of structural states of the system into the set of
    qualitative states of the system $Q$.
    \begin{equation}\label{eq:structural-qualities}
    \phi : S \to Q
    \end{equation}
    Note that the dynamics of the system can be fully recovered via (\ref{eq:structural-dynamics}), without reference to
    $Q$.
    This can be seen as an epiphenomenal picture, in which the qualitative states $Q$ are recognized by the mapping
    $\phi$, but have no influence over the system's evolution in time.

    \subsubsection{Qualitatively-Determined Dynamics}\label{subsubsec:qualitatively-determined-dynamics}

    In the case of a qualitatively-determined system, the dynamics are wholly independent of the structure of the system
    $S$; they can be stated as the dual of (\ref{eq:structural-dynamics}).
    \begin{equation}\label{eq:qualitatively-determined-dynamics}
    F_i : Q \times T \to Q.
    \end{equation}
    In these cases, there may or may not be a mapping $\psi$ such that some structure can be recovered.
    \begin{equation}\label{eq:quality-to-structure}
    \psi : Q \to S
    \end{equation}

    \subsubsection{Mixed-Determination Dynamics}\label{subsubsec:mixed-determination-dynamics}

    Now we can see that in qualitatively-determined systems, (\ref{eq:structural-dynamics}) is insufficient to determine
    the behavior of the system; we require an explicit reference to $Q$.
    In this case, our $F_i$ become
    \begin{equation}\label{eq:intrinsic-dynamics}
    F_i : S \times Q \times T \to S \times Q.
    \end{equation}
    This is the case of a mixed-determination system.

    Included in this category are also certain neutral monist systems, which do not fall into the equivalent category below.
    In this picture, the fundamental dynamics are defined a set of fundamentally \emph{neutral} entities $N$, which are
    neither exclusively qualitative nor exclusively structural.
    In this specific mixed-determination monistic view, the structural and qualitative properties are determined by the neutral ones,
    but the neutral ones are not fully-determined by the structural and qualitative ones

    We can formulate this as another monistic dynamics, i.e.\ analogous to (\ref{eq:structural-dynamics}) and (\ref{eq:qualitatively-determined-dynamics}):
    \begin{equation}\label{eq:neutral-dynamics}
    F_i : N \times T \to N.
    \end{equation}
    In order for this picture to be meaningfully different from (\ref{eq:intrinsic-dynamics}) and (\ref{eq:qualitatively-determined-dynamics}),
    there can be no deterministic mapping from $S$ and $Q$ to recover the neutral entities $N$, that is
    \begin{equation}\label{eq:no-neutral-mapping}
    \nexists\, \nu \colon S \times Q \to N.
    \end{equation}
    However, there \emph{must} be mappings $\phi$ and $\psi$ such that
    \begin{equation}\label{eq:neutral-mapping}
    \phi : N \to Q, \quad \psi : N \to S.
    \end{equation}

    \subsubsection{Equivalent Dynamics}\label{subsubsec:equivalent-dynamics}

    In the equivalent example we gave, there were two alternative definitions of the behavior of the system.
    In the structural dynamics, each state in $S$ is a matrix $s$ whose columns and rows represent the cells on the grid.
    Each $s_{ij}$ is 1 when active and 0 when inactive.

    In the Game of Life, the transition rules are defined only by the neighborhood around a cell, some $\tilde{s}_{uv}$.
    We then take $u = \{i-1, i, i+1\}$ and $v = \{j-1, j, j+1\}$, and call the set of all states of this neighborhood
    $\tilde{S}$.

    The evolution of the system then takes place according to the analog of (\ref{eq:structural-dynamics}),
    \begin{equation}\label{eq:equiv-structural-dynamics}
    F_i : \tilde{S} \times T \to \tilde{S}.
    \end{equation}
    Note now that we have an analog of (\ref{eq:structural-qualities}) that maps between these neighborhoods and their
    encoding into the set of colors $Q$; the specific state of each neighborhood is mapped to exactly one color.
    \begin{equation}\label{eq:equiv-structural-qualities}
    \phi : \tilde{S} \to Q
    \end{equation}

    And in turn we can state the dynamics in an alternate form $G$, entirely in terms of $Q$:
    \begin{equation}\label{eq:equiv-intrinsic-dynamics}
    G_i : Q \times T \to Q,
    \end{equation}
    which can be reversed by our analog of (\ref{eq:quality-to-structure}):
    \begin{equation}\label{eq:equiv-quality-to-structure}
    \psi : Q \to \tilde{S}.
    \end{equation}

    We can see the commutative nature of this arrangement.
    \begin{gather}
        F_i : \psi(Q) \times T \to \psi(Q)\\
        G_i : \phi(\tilde{S}) \times T \to \phi(\tilde{S})
    \end{gather}

    If we consider the $F_i$ as morphisms of a category $\mathcal{F}$, with the $\tilde{S}$ and $T$ as objects, then the
    $G_i$ are morphisms of a category $\mathcal{G}$ with the $Q$ and $T$ as objects.\footnote{For more information on
    applied category theory see \cite{spivak2014}.}
    We can then generalize the $\phi$ and $\psi$ as functors.
    \begin{equation}\label{eq:functors}
        \phi: \mathcal{F} \to \mathcal{G}, \quad \psi: \mathcal{G} \to \mathcal{F}
    \end{equation}

    And we can see their commutative relationship as follows:
    \[
        \begin{tikzcd}[row sep=large, column sep=huge]
            \tilde{S} \times T \arrow[r, "F_i"] \arrow[d, "\phi", shift left=1] & \tilde{S} \arrow[d, "\phi", shift left=1] \\
            Q \times T \arrow[r, "G_i"'] \arrow[u, "\psi", shift left=1] & Q \arrow[u, "\psi", shift left=1]
        \end{tikzcd}
    \]

    We can now see that the dynamics defined by $F$ and $G$ share an equivalence:
    \begin{gather}
        \phi \circ F_i = G_i \circ \phi, \\
        \psi \circ G_i = F_i \circ \psi.
    \end{gather}
    The potential significance of the relationship will be seen as we proceed.

    Although the example we gave in the illustrations section feels a bit contrived, it would be interesting to
    find non-trivial equivalent mappings.
    An equivalence between a structural system $F_i : S \times T \to S$ and a mixed-determination one
    $G_i : S' \times Q \times T \to S' \times Q$ could also be fruitful to investigate.

    The commutative relationship brings to mind the equivalence between Newtonian and Hamiltonian-Lagrangian mechanics, where the
    Newtonian dynamics are defined as a set of forces on a configuration space, whose points are particles encoded with a
    position at some time $t$, while the equivalent Hamiltonian-Lagrangian dynamics is defined as an extremum of some
    functional on a phase space, whose points encode positions and momenta.

    I suspect that the property of equivalence, if not an entirely generic feature of continuous systems, is much easier to encounter
    within the continuum that in the discrete world of cellular automata.

    \subsection{Informatics}\label{subsec:informatics}

    In addition to dynamics, we can analyze our categories of models according to their informational properties.
    As we look at this picture, we realize that each category represents its own non-trivial informational relationships
    between structural and qualitative properties.

    \begin{enumerate}
    \item In the functionalist case, there is simply no correlation between the structural properties and the qualitative
    ones (usually none).
    \item In the structural case, qualitative properties can be derived from structural ones, but it may not always be the case in the reverse.
    \item In the qualitatively-determined case, structural properties may be derived from qualitative ones, but not necessarily in the reverse.
    \item In the mixed-determination case, structural and qualitative properties are derived from structural and qualitative properties:
    they are both required.
    \item In the fully-equivalent case, there is perfect correlation between the structural properties and the qualitative ones;
    either can be completely derived from information about the other.
    \end{enumerate}

    \subsubsection{Q-S Space}\label{subsubsec:qs-space}

    \emph{Q-S Space} is a 2-dimensional representation of the amount of information the structural and qualitative
    spaces determine about each other.
    Any system can be represented as a pair $(\hat{q},\hat{s})$.
    The first coordinate $\hat{q}$ is the level of certainty in structure given quality, the second $\hat{s}$ is the certainty in quality given structure.
    1 implies perfect certainty, while 0 indicates no correlation.

    Each coordinate can be calculated as follows:
    \begin{equation} \label{eq:q-hat}
        \hat{q} = Z(\text{Structure} \mid \text{Quality}),
    \end{equation}
    \begin{equation}\label{eq:s-hat}
        \hat{s} = Z(\text{Quality} \mid \text{Structure}).
    \end{equation}
    Intuitively, $Z(A \mid B)$ is the amount of uncertainty in $A$ explained by $B$, normalized in the range $[0,1]$.
    
    In the discrete case, we define $Z$ in relation to the normalized conditional entropy.
    \begin{equation}\label{eq:c-discrete}
    Z(A \mid B) = \frac{H(A \mid B)}{H(A)}
    \end{equation}
    Here $H$ is the conventional information-theoretic entropy.

    In continuous domains, the previous may no longer be bound within the interval $[0,1]$.
    In these cases we may instead define $Z(A \mid B)$ as the correlation ratio $\eta$.
    \begin{equation}\label{eq:c-continuous}
    Z(A \mid B) = \eta_{a \mid b} = \sqrt{\frac{\Var(\expectation[A \mid B])}{\mathrm{Var(A)}}}
    \end{equation}
    Its essential meaning is extremely similar; $Z(A \mid B)$ gives the ratio of variation in our expectation
    of $A$ explained by $B$.

    \begin{figure}[htbp]
        \centering
        \begin{tikzpicture}[scale=4, font=\small]

            \draw[->] (0,0) -- (1.1,0);
            \draw[->] (0,0) -- (0,1.1);

            \node[rotate=0, anchor=north, font=\small] at (0.5, -0.02) {$\hat{q}$};
            \node[rotate=90, anchor=south, font=\small] at (-0.02, 0.5) {$\hat{s}$};

            \draw[dashed, gray] (1,0) -- (1,1);
            \draw[dashed, gray] (0,1) -- (1,1);

            \node[circle, draw, inner sep=1.2pt, line width=1pt, label=below left:{\textbf{Functional} (0,0)}] at (0,0) {};
            \node[circle, draw, inner sep=1.2pt, line width=1pt, label=above left:{\textbf{Structural} ($\hat{q}$,1)}] at (0,1) {};
            \node[circle, draw, inner sep=1.2pt, line width=1pt, label=below right:{(1,$\hat{s}$) \textbf{Qualitative}}] at (1,0) {};
            \node[circle, draw, inner sep=1.2pt, line width=1pt, label=above right:{(1,1) \textbf{Equivalent}}] at (1,1) {};
            \node[circle, draw, inner sep=1.2pt, line width=1pt, label=above:{\textbf{Mixed}}] at (0.5,0.5) {};

        \end{tikzpicture}
        \caption{Five model types represented as $(\hat{q},\hat{s})$ pairs in \emph{Q-S space}.
        Points that lie on the dashed lines (where either or both coordinates are 1) are referred to as \emph{fully-determinate}.
        Either their quality or their structure (or both) is sufficient to describe the entire system, so long as the appropriate
        derivation of the other is known.}
        \label{fig:entropy-map}
    \end{figure}
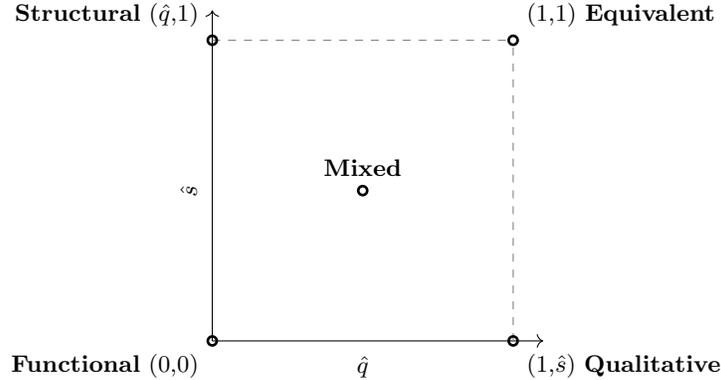

    \begin{figure}[htbp]
        \centering
        \includegraphics[width=0.65\textwidth]{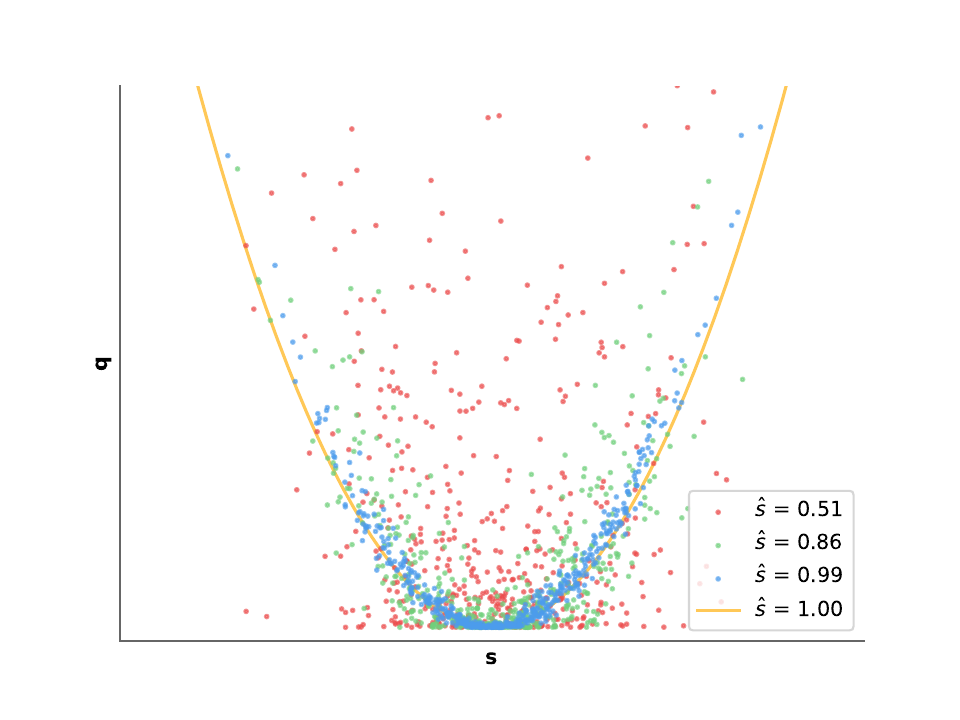}
        \caption{
            Scatterplot illustrating the fidelity between $q$ and $s$ for different values of $\hat{s}$.
            This data was derived from simulations which modeled $\phi(s)$ as approximately quadratic, with the result that
            a perfectly correlated $q$ would follow a parabolic curve.
            Noise was injected to model different levels of correlation between $q$ and $s$.
        }
        \label{fig:q_s_fidelity}
    \end{figure}

    \subsubsection{Calculating Informatics}\label{subsubsec:calculating-informatics}

    In the case of structurally-determined systems, the association between structural and qualitative properties is
    entirely set by (\ref{eq:structural-qualities}).
    Assuming the relationship is deterministic, it means that our structural properties fully determine the qualitative
    ones, i.e. $\hat{s}=1$.
    The reverse case is dependent on the exact nature of our implementation of $\phi$.

    Suppose $S = \{0, 1, 2\}$, $Q = \{0, 1\}$, and
    \begin{gather}\label{eq:first-q}
    q = \phi(s) = \begin{cases}
               1 & \text{if } s = 0, \\
               1 & \text{if } s = 1, \\
               0 & \text{if } s = 2.
    \end{cases}
    \end{gather}
    Then with some calculation we find that ${\hat{q} \approx 0.579}$.

    These techniques can be extended to take into account differences over time through the dynamics.
    For examples in discrete-time systems, my previous work\footnote{\cite{williams2024-grounded}} discusses divergence 
    metrics that can be adapted for this purpose.
    Analytic equivalents can be constructed for continuous-time systems, with techniques varying based on the
    formulation of the dynamics.

    \subsection{Interpretation}\label{subsec:interpretation}

    Now is a good time to take a step back towards the concrete so we can interpret the meaning of our categories.

    \subsubsection{Bare Functionalism}\label{subsubsec:bare-functionalism}

    A functionalist system does not seem to have any basis for a correlation between qualitative properties and
    structural properties that would be needed for natural selection to prefer certain qualitative states.
    This would not likely bother the functionalist much, but it is most likely that they reject the existence of the
    qualitative altogether.
    The alternative viewpoint appears some to be a type of dualism with a pair of utterly disconnected worlds.

    \subsubsection{Structurally-Determined Systems}\label{subsubsec:structurally-determined-systems}

    The structuralist view is one that seems most commonly assumed in the sciences.
    It can be found in very prominent theories including IIT\footnote{\cite{tononi2016}}, Global Workspace Theory\footnote{\cite{baars2005}, \cite{dehaene2011}}, and attention schema theory\footnote{\cite{graziano2015}, \cite{graziano2019}}.

    The dynamics of the world are described fully by physics, which exhibit a form of causal closure.\footnote{\cite{Carroll2021}}
    The addition of the qualitative is then wholly epiphenomenal.\footnote{\cite{Jackson1982}}
    It does not seem to imply any causation from the qualitative to the structural, though we would expect in most cases
    $\hat{q} > 0$.
    These correlations may be seen as accidental, or at least incidental, since the dynamics can be entirely described without them.

    A meaningful distinction can be drawn between emergentist views and panpsychist ones.
    In the emergentist view, qualitative properties are associated with only specific states; in other words that certain structural states map to no qualitative state:
    \begin{equation}\label{eq:emergentism}
        \exists s \in S : \phi(s) = \emptyset.
    \end{equation}
    From the panpsychist side, every structural state has a qualitative one:
    \begin{equation}\label{eq:panpsychism}
    \forall s \in S, \ \phi(s) \in Q.
    \end{equation}

    \subsubsection{Mixed-Determination Systems}

    Mixed-determination systems are subtle to piece apart.
    Ultimately, if equation (\ref{eq:intrinsic-dynamics}) cannot be decomposed into an equivalent system where some
    ${F_i : S \times T \to S}$ and ${G_i : Q \times T \to Q}$, it means that ${S \times Q}$ cannot be treated
    independently.

    This picture aligns with a dualist view; in some way the state of the world must be stated
    all at once with its qualitative properties along with its structural ones.
    The qualitative properties are intimately baked into the fundaments of the world in such a way that a change to
    the qualitative state is a change in the whole state; its behavior will change.
    This fits with our intuitive notions of mental events causing and being caused by physical events, and represents a
    clear break from causal closure.
    This would in principle be empirically observable.

    An alternative consideration of mixed-determination systems includes the neutral system we described in (\ref{eq:neutral-dynamics}).
    In this picture, the structural and qualitative are both shadows of an underlying neutral world that may
    be only partially manifest in either domain.

    While $(\hat{q}>0,\hat{s}>0)$, neither can fully determined by the other $(\hat{q} \neq 1,\hat{s} \neq 1)$, but is
    instead derived from the underlying $N$.

    We can imagine a generalization of \emph{Q-S space}, whose coordinates are
    \begin{equation}\label{eq:qs-generalized}
    (\hat{q}, \hat{s}, \hat{q_n}, \hat{s_n}) = Z(S \mid Q)
        \,\times\, Z(Q \mid S)
        \,\times\, Z(N \mid Q)
        \,\times\, Z(N \mid S)
    \end{equation}
    where it is understood that
    \begin{equation}\label{eq:N}
    Z(S \mid N) = Z(Q \mid N) = 1
    \end{equation}
    by hypothesis.

    \subsubsection{Qualitatively-Determined Systems}

    Qualitatively-determined systems may be seen as idealist.
    The structural properties of the world are derivative of the qualitative state, the inverse of the structural
    picture.
    Because of that duality, almost anything that can be stated about a structuralist picture can be stated in reverse
    about the qualitatively-determined.
    The differences and similarities between those two views merits substantial thought on its own that cannot
    be fully explored in this paper.

    However, the intuitively most obvious objection to this type of picture is the apparent causal closure of
    the physical.
    If our idea is that the physical is the structural, then if both the physical system has causal closure \emph{and}
    the qualitative system has causal closure (by hypothesis as a qualitatively-determined system), then the world would by
    definition be an equivalent system, not a qualitatively-determined one.

    The most obvious way to parry this objection is to appeal to quantum mechanics, positing that the structural,
    by way of incorporating inherently non-deterministic dynamics (on some interpretations at least), is an
    incomplete picture of the system, whose dynamics are not fully-determined in $S$ but rather in $Q$.
    This picture in \emph{Q-S space} yields high determination in $S$ but perfect determination in $Q$,
    ${(\hat{q} = 1, \hat{s} \gg 0)}$.\footnote{This view is consistent with \cite{kastrup2019idea}.}

    I'm not qualified to remark on whether this defense is valid, but on conceptual grounds it seems plausible.
    That of course does not provide any direct empirical evidence for a qualitatively-determined picture, but it could
    remove a potentially fatal objection.
    It should be noted that even if QM could be verified as non-deterministic, it's entirely possible that the
    dynamics of $S$ are just inherently non-deterministic, with no further determinism to be recovered in $Q$ or anywhere else.

    \subsubsection{Equivalent Systems}

    The equivalent systems are the ones I find most fascinating.
    An equivalent system is one that can be equally described according to two sets of dynamical rules, either of
    which constitutes a full and equivalent picture of the system.
    This arrangement summons to mind varieties of neutral monism, especially Russellian monism\footnote{\cite{russell1921}, \cite{russell1927}}.

    On Russell's picture, the world can be viewed as a directed graph $U = (N,R)$ consisting of a set $N$ of neutral
    entities and $R$, a set of ordered pairs $(v \to w)$ representing interactions between entities, where
    $v, w \in N$.\footnote{For a full exploration of metaphysical systems in this manner see \cite{williams2024}.}
    It can be natural to understand each pair to be an interaction in which $v$ is the cause and $w$ the effect.

    \begin{figure}[thbp]
        \centering
        \begin{subfigure}[b]{0.33\textwidth}
            \centering
            \includegraphics[width=\textwidth]{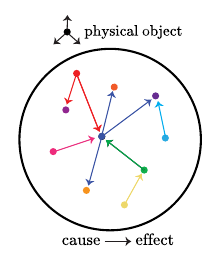}
            \caption{Physical perspective}
            \label{fig:russell_physical}
        \end{subfigure}
        \begin{subfigure}[b]{0.33\textwidth}
            \centering
            \includegraphics[width=\textwidth]{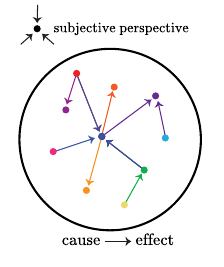}
            \caption{Subjective perspective}
            \label{fig:russell_mental}
        \end{subfigure}
        \caption{The graph structure for Russellian monism. The left panel emphasizes the physical aspect by color coding groups of outgoing arrows based on their source, while the right panel highlights the mental aspect by grouping arrows according to their destination.}
        \label{fig:russell_monism}
    \end{figure}

    On Russell's view, there are two different but ultimately equivalent ways to view this system.
    The first is as a set of sets of ordered pairs $V$, each grouped by a common cause $V$.
    In this way, each event is grouped together with its cause; each set is a collection of the rays extending outward
    from its source.
    Inversely, the second is as a set of sets of ordered pairs $W$, each grouped by a common effect $w$.
    In this picture, each ray is grouped together with the others it points in towards.

    We can see that the underlying graph $U$ is equally represented if it is parsed out as either of the sets of
    sets $V$ or $W$.
    We can define our structural state to be the set $V \in S$; we take this as the view of objects projecting their
    influence on each other.

    And we can similarly define our qualitative state to be the set $W \in Q$; we take this as a set of perspectives,
    each collecting the experiential contents emanating towards a single point.\footnote{This symmetry brings to mind
    the duality of $\text{Hom}(A, -)$ and $\text{Hom}(-,A)$ in the Yoneda Lemma, which demonstrates that
    an object can be determined by either its morphisms into or out of other objects; see \cite{maclane1971categories}.}

    Now if we can decompose our dynamics ${F_i : S \times T \to S}$ and ${G_i : Q \times T \to Q}$ in a way that yields
    the same behavior of the underlying graph $U$, we have a wholly equivalent system.

    This can be illustrated in a system where each node $n_i \in N$ of the graph has an associated value $x_i \in \mathbb{R}$.
    We can then express the dynamics on this system in two different but equivalent ways, similar in nature to the
    equivalence between Newtonian and Hamiltonian-Lagrangian systems in classical mechanics.

    The first approach involves defining an explicit set of differential equations that specify the evolution of each node
    individually with respect to each of its afferent neighbors (i.e.\ those for whom it has an edge $(v \to w) \in R$).

    This is akin to the Newtonian formulation, in which each force is defined explicitly.

    We can state these as a set of differential equations indicating the flow with respect to each $x_i$
    \begin{equation}\label{eq:equiv-differential}
    \frac{dx_i}{dt} = \beta \sum_{W(n_i)} (x_j - x_i),
    \end{equation}
    where $W(n_i)$ is the set of edges afferent to $n_i$
    \begin{equation}\label{eq:afferent}
    W(n_i) = \{ (n_j \to n_i) \in R \}.
    \end{equation}
    Because $W \in Q$, this view can be seen as the qualitative view.

    The alternative approach is to define an energy function for the entire system, which the dynamics will act to
    keep stable.
    \begin{equation}\label{eq:equiv-energy}
    E(x) = \frac{1}{2} \sum_{n_i \in N} \sum_{V(n_i)}(x_j - x_i)^2
    \end{equation}
    where $V(n_i)$ is the set of edges efferent to $n_i$
    \begin{equation}\label{eq:efferent}
    V(n_i) = \{ (n_i \to n_j) \in R \}.
    \end{equation}
    Because $V \in S$, this view can be seen as the structural view.
    It should be apparent the parallels this approach has to the Hamiltonian-Lagrangian formulation of classical mechanics.

    Since addition is commutative, we can change (\ref{eq:equiv-energy}) to sum over the efferent edges:
    \begin{equation}\label{eq:equiv-energy-efferent}
    E(x) = \frac{1}{2} \sum_{n_i \in N} \sum_{W(n_i)}(x_j - x_i)^2.
    \end{equation}
    Then we can calculate the gradient of this function with respect to $x_i$ as
    \begin{equation}\label{eq:equiv-gradient}
    \frac{\partial E}{\partial x_i} = \sum_{W(n_i)} (x_j - x_i),
    \end{equation}
    and the gradient flow as
    \begin{equation}\label{eq:equiv-gradient-flow}
    \frac{dx_i}{dt} = \gamma \sum_{W(n_i)} (x_j - x_i).
    \end{equation}
    Note the equivalence between this form and the earlier (\ref{eq:equiv-differential}).

    I find a few points of attraction in the equivalent view.
    Admittedly, none of these is a solid case that the picture is right; but they entice me to want to look closer.

    \begin{enumerate}
        \item First, it rings out to me with a certain beauty that we often find in truth.

        \item Second, the ability to decompose the dynamics equivalently is not a generic feature of all systems; it should
    substantially narrow down the search space of hypotheses for these types of relations.
    And while being the drunk man looking under the lamp post for his keys isn't ideal, it is the most
    rational place to look first.

     \item The equivalent view also seems to open the door to the possibility that we might be able to give a principled (or at least
    partly-principled) description of a certain set of phenomena; it might be provably the most parsimonious description
    that exactly accounts for the two alternative dynamics.

     \item Finally, there is a certain elegance in its implications for mental-physical causation and vice versa; it gives
    both mental-physical causation and physical-mental causation without any additional machinery.
    It is just the nature of an equivalent system to have a coherence between the two perspectives.
    \end{enumerate}

    \subsection{Applications}\label{subsec:applications}

    The \emph{Q-S systems} described in this paper have already been applied to derive formal constraints on the evolution
    of consciousness in the companion article \emph{Qualia \& Natural Selection}.
    It could serve as a framework for further work in evolutionary biology, quantitative neuroscience, and cognitive science.

    \pagebreak

    \bibliographystyle{apalike}
    \bibliography{references}

\end{document}